# How Temporal Dispersion of Radio Channels Depends on Signal Bandwidth

Jarosław Wojtuń[1], Tomasz Grasza[1], Cezary Ziółkowski[1], Jan M. Kelner[1], Aniruddha Chandra[2], Radek Zavorka[3], Josef Vychodil[3], Jiri Blumenstein[3], Tomas Mikulasek[3], Ales Prokes[3]

[1] Institute of Communications Systems, Faculty of Electronics, Military University of Technology, Warsaw, Poland, *jaroslaw.wojtun@wat.edu.pl*

[2] ECE Department, National Institute of Technology, Durgapur, India

[3] Department of Radio Electronics, Brno University of Technology, Brno, Czech Republic

***Abstract*—The degree of temporal dispersion of a radio channel is the basis for assessing its transmission capabilities. This paper addresses the practical issue of measuring the temporal parameters of a signal transmitted in the millimeter-wave (38 GHz) range. These parameters provide the basis for assessing the degree of dispersion of the received signal. The measurement results show that both the mean and RMS values of the delay spread depend on the distance between the transmitter and receiver as well as the bandwidth of the transmitted signals. The function description of these relationships (model) is determined based on a statistical analysis of the obtained measurement data obtained in a selected indoor corridor under line-of-sight conditions.**

## I. Introduction

The dynamic growth in demand for mobile telecommunications services has led to an expansion of the spectral resources allocated to individual communication channels. Consequently, the operating frequency bands of radio devices are being shifted toward the millimeter-wave (mmWave) range, as exemplified by the development of fifth-generation (5G) and future wireless technologies [1].

Accurate assessment of the transmission characteristics of different environments and propagation conditions is essential for the design of efficient transmission structures and signal processing methods operating in these higher frequency bands. A thorough understanding of the actual propagation parameters of radio channels requires extensive measurement campaigns conducted across various environments and under diverse atmospheric conditions. Numerous publications, such as [2], [3], [4], [5], [6], [7], demonstrate the breadth of research carried out in this area.

Early studies on mmWave propagation primarily focused on developing path loss (PL) models that consider antenna radiation patterns and environmental types. Examples of such analyses can be found in [4], [7]. However, current research goes beyond simple PL characterization. It includes the analysis of temporal and angular characteristics of the received power distribution in the vicinity of the receiving antenna. Among the numerous studies whose results form the basis for developing propagation models in terms of signal level, time delay, delay spread, angle of arrival (AOA), and angle of departure (AOD), the works presented in [2], [3], [5], [6] deserve particular attention.

Measurements and evaluation of temporal parameters such as mean delay and root mean square (RMS) delay spread are fundamental for assessing the transmission capabilities of radio channels. These parameters determine the achievable information rate in a given radio link. The mentioned works present primarily models in the table form or graphs of statistical parameters (such as the RMS delay spread) derived from measurement data that primarily account for environmental type and propagation conditions.

Measurement results obtained by various research groups have served as the foundation for developing statistical models describing the influence of propagation on the temporal characteristics of received signals. One such model, presented in [8], enables the generation of parameters (such as power, AOA, AOD, and time delay) for individual multipath components reaching a receiver (Rx) via different propagation paths. In this model, environmental conditions, distance, and frequency are incorporated through scaling factors.

In contrast, the present paper aims to emphasize the factors that directly affect the temporal characteristics of the received signal. Specifically, we analyze not only the impact of propagation conditions and environment type but also the effects of changes in the distance between the transmitter (Tx) and Rx, as well as the influence of the transmitted signal bandwidth (BW). Measurements were conducted at 38 GHz for a selected indoor propagation scenario. The obtained results make it possible to demonstrate how variations in the Tx–Rx distance and signal BW influence the temporal dispersion of the received signal.

Based on the analysis of measurement data, empirical models describing the variation of these parameters as functions of the Tx–Rx distance and signal BW are proposed. Although the presented results are based on a specific propagation scenario, they may serve as a foundation for developing a broader statistical model that integrates measurement results from various research centers and scenarios. These preliminary findings highlight the need for further studies in this direction and underscore the originality of the results presented in this work.

This research was funded in part by the National Science Center (NCN), Poland, grant no. 2021/43/I/ST7/03294 (MubaMilWave).

The remainder of the paper is organized as follows. Section II introduces the phenomenon of radio channel dispersion, focusing on temporal ones. Whereas Section III describes the applied channel sounder and spatial scenario. Section IV discusses the measurement results and provides estimation outcomes demonstrating the relationships between temporal dispersion parameters, Tx–Rx distance, and transmitted signal BW. Finally, Section V concludes the paper.

## II. Channel Dispersion

Radio wave propagation in realistic environments is inherently affected by multipath propagation, which arises when transmitted signals reach the Rx through multiple paths of different lengths. These paths are formed due to reflections, diffractions, and scattering caused by environmental features such as buildings, vegetation, vehicles, technical infrastructure, and even moving vehicles or people. As a result, the received signal is a superposition of multiple components arriving at different delays, angles, and often with varying phases and amplitudes. The phenomenon of multipath propagation leads to dispersion of the received signal, i.e., the spreading of its energy over time, frequency, or AOA domain, which directly affects the transmission quality and determines the achievable data rate. The characteristics of channel dispersion strongly depend on the anisotropy of the propagation environment, resulting from its geometric and material inhomogeneity, and the presence of both static and dynamic scattering objects.

Depending on the domain of observation, dispersion effects can be categorized as:

- time-domain dispersion, resulting from differences in propagation path lengths and corresponding arrival times of multipath components;
- frequency-domain dispersion, which is manifested as selective fading of certain frequency components within the transmitted signal band;
- angular dispersion, describing the spread of arriving signal components over different AOA and AOD.

In dynamic environments, Doppler dispersion additionally appears in the frequency domain, caused by the relative motion of the Tx, Rx, or moving scatterers that alter the apparent frequency of the received signal components.

To quantify and model the temporal dispersion of a radio channel, the power delay profile (PDP) is commonly used. The PDP represents the distribution of the received power as a function of delay and provides a statistical description of the multipath structure. From the PDP, two basic parameters are derived: the mean excess delay and the RMS delay spread. These parameters reflect, respectively, the average delay of received signal components and the dispersion of their arrival times around the mean delay. In particular, the RMS delay spread is a key indicator of the temporal selectivity of the channel.

The measured PDP typically has a continuous form, $P(\tau)$. However, for modeling purposes, it is often represented in a discrete form, $P(\tau_k)$, e.g., tapped-delay line (TDL), where each tap corresponds to a distinct multipath component characterized by its power (or attenuation) and delay. In practical terms, the discrete parameters are determined based on the local extrema (peaks) of the continuous PDP function, which correspond to the dominant multipath components observed in the received signal. In such a representation, the PDP is expressed as a set of discrete values, $\{\tau_k, P_k\}_{k=1,2,\ldots,K}$, where $P_k = P(\tau_k)$ denotes the average power of the received component with propagation delay $\tau_k$.

Based on continuous and discrete PDPs, the mean excess delay $\bar{\tau}$ and the RMS delay spread $\sigma_\tau$ are defined as follows:

$$\begin{aligned} P(\tau)&: \quad \bar{\tau} = \frac{1}{\bar{P}}\int_0^\infty P(\tau)\tau d\tau\,, \\ P(\tau_k)&: \quad \bar{\tau} = \frac{1}{\bar{P}}\sum_{k=1}^{K} P_k\tau_k\,, \end{aligned} \tag{1}$$

$$\begin{aligned} P(\tau)&: \quad \sigma_\tau = \frac{1}{\bar{P}}\sqrt{\int_0^\infty P(\tau)(\tau-\bar{\tau})^2 d\tau}\,, \\ P(\tau_k)&: \quad \sigma_\tau = \frac{1}{\bar{P}}\sqrt{\sum_{k=1}^{K} P_k(\tau_k-\bar{\tau})^2}\,, \end{aligned} \tag{2}$$

where $\bar{P}$ is the average received power:

$$\begin{aligned} P(\tau)&: \quad \bar{P} = \int_0^\infty P(\tau) d\tau\,, \\ P(\tau_k)&: \quad \bar{P} = \sum_{k=1}^{K} P_k\,. \end{aligned} \tag{3}$$

A larger value of $\sigma_\tau$ indicates a higher degree of temporal dispersion in the propagation channel, leading to a greater potential for intersymbol interference (ISI) in broadband communication systems. Consequently, precise estimation of PDP and its parameters is essential for designing and evaluating high-data-rate radio links, particularly in mmWave communication systems.

## III. Channel Sounder and Measurement Scenario

### *A. Measurement Setup*

The measurements were carried out using a wideband channel sounder designed for propagation studies in the mmWave frequency range. The core of the measurement system consisted of a Keysight M9384B VXG microwave vector signal generator on the transmitting side (Tx) and a Keysight N9032B PXA signal analyzer on the receiving side (Rx). The signal generator operated over a frequency range from 1 MHz to 44 GHz, while the signal analyzer covered a range from 2 Hz to 44 GHz, both supporting a measurement BW of up to 1 GHz.

The Tx included the signal generator and a transmitting antenna connected through a low-loss feeder. The Rx comprised the receiving antenna, feeder, signal analyzer, and a computer unit responsible for data acquisition and signal recording. To maintain high time and frequency stability, the generator and analyzer were synchronized using a Stanford Research FS725 rubidium frequency standard.

An Eravant (formerly SAGE Millimeter, Inc.) SAR-2507-28-S2 rectangular horn antenna was used on both the Tx and Rx sides. The antenna operates over the 26.5–40 GHz frequency range, featuring half-power beamwidths (HPBWs) of 7° and 9° in the elevation and azimuth planes, respectively, and provides a gain of 25 dBi.

A frequency-modulated continuous-wave (FMCW) chirp signal was used as the sounding waveform. Two FMCW waveforms with BWs of 100 and 900 MHz were employed to enable high-resolution characterization of the multipath propagation profile and to analyze BW-dependent effects. Each FMCW pulse lasted 8 µs, followed by a 2 µs idle interval. The sampling frequency of both the Tx and Rx chains was set to 1.2 GHz, with the observed frequency difference determined by the BW of the sounding signal. On the Rx side, a total of 10 ms of data were recorded, corresponding to 1,000 sounding pulses captured per measurement position. All measurements were conducted at a carrier frequency of 38 GHz. A schematic representation of the test-bed and scenario is shown in Fig. 1 [9].

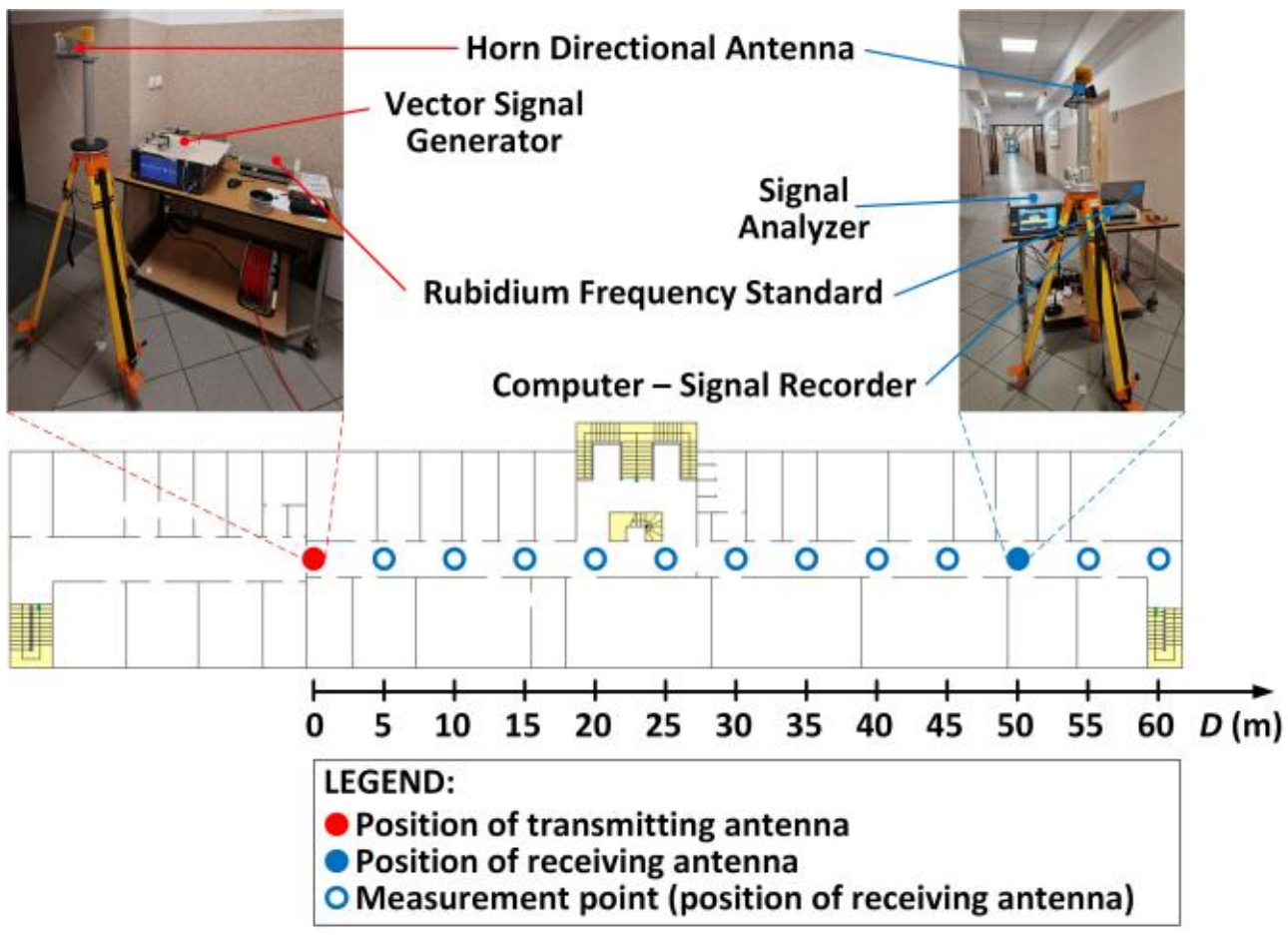

Fig. 1. Measurement scenario and setup configuration.

### B. Measurement Scenario

The measurement campaign was conducted inside a two-story Institute of Communication Systems building at the Faculty of Electronics, Military University of Technology (WAT). The tests were performed on the second-floor corridor, representing a typical indoor line-of-sight (LOS) propagation environment.

The Tx part of the setup was installed at one end of the corridor, just before its narrowing, to minimize additional reflections and propagation constraints. The Rx unit was successively repositioned along the corridor axis, forming a set of twelve Rx locations separated by 5 m intervals. The measurement points were thus distributed at distances ranging from 5 to 60 m from the Tx antenna.

This configuration allowed for evaluating the impact of the sounding signal BW and distance on the channel temporal dispersion characteristics.

## IV. Evaluation of Channel Temporal Dispersion

According to the scenario described in Section III, measurements of the channel transmission characteristics were performed for twelve Rx locations distributed along the corridor and for two analyzed signal bandwidths, 100 and 900 MHz.

At each measurement point and for each bandwidth, 10 ms of signal data were recorded, corresponding to 1,000 FMCW sounding pulses. For every selected Tx–Rx distance, the received pulses were averaged to obtain a representative waveform, which was the basis for determining the averaged PDP normalized to the power and delay of the direct-path component.

Figs. 2–4 present representative examples of the averaged PDPs obtained for two analyzed signal BWs (100 and 900 MHz are marked in red and blue, respectively) and three selected Tx–Rx separation distances, i.e., 10, 30, and 50 m, respectively.

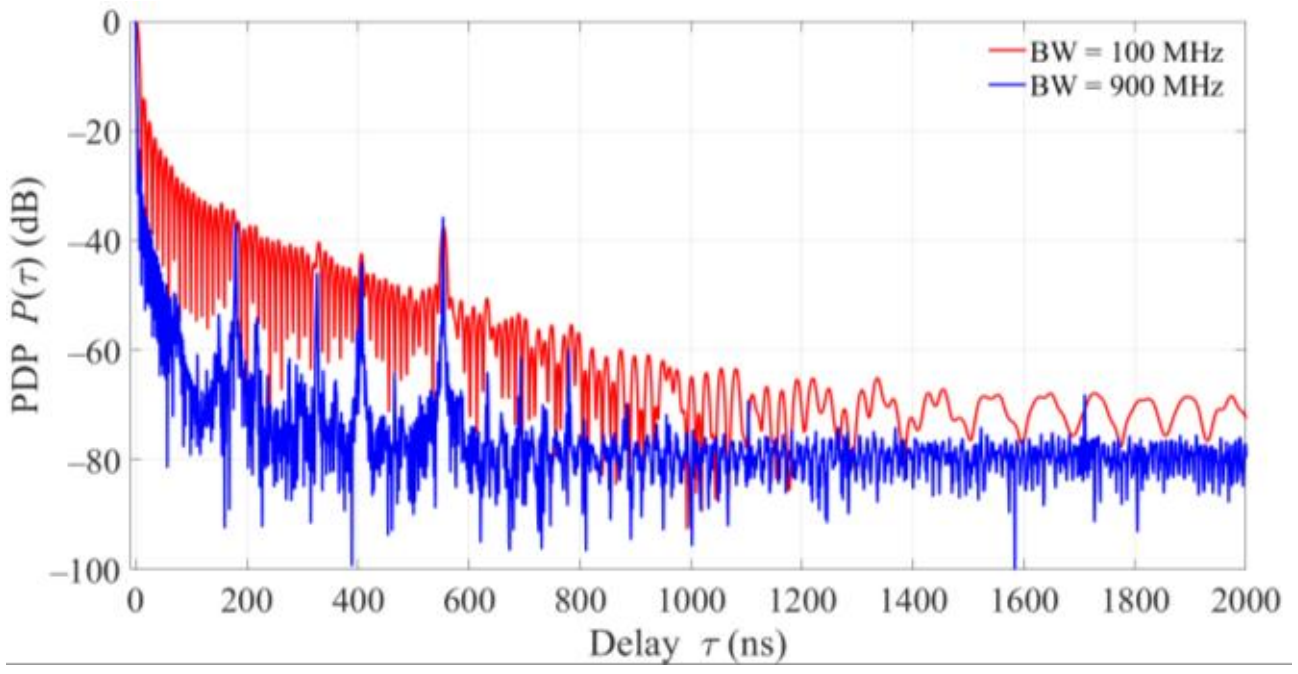

Fig. 2. Averaged PDPs for 100 and 900 MHz BWs, carrier frequency of 38 GHz, and at Tx–Rx distance of 10 m.

At the shortest distance, i.e., 10 m (see Fig. 2), the PDP exhibits a clearly dominant first component corresponding to the LOS path, followed by a number of weaker delayed components. The difference between the two BWs is particularly visible in the initial portion of the PDP: For 900 MHz, waveform provides noticeably finer temporal resolution, which allows the detection of more distinct multipath components within the same delay window. The narrower 100 MHz signal results in a smoother, more averaged power distribution due to its lower time resolution.

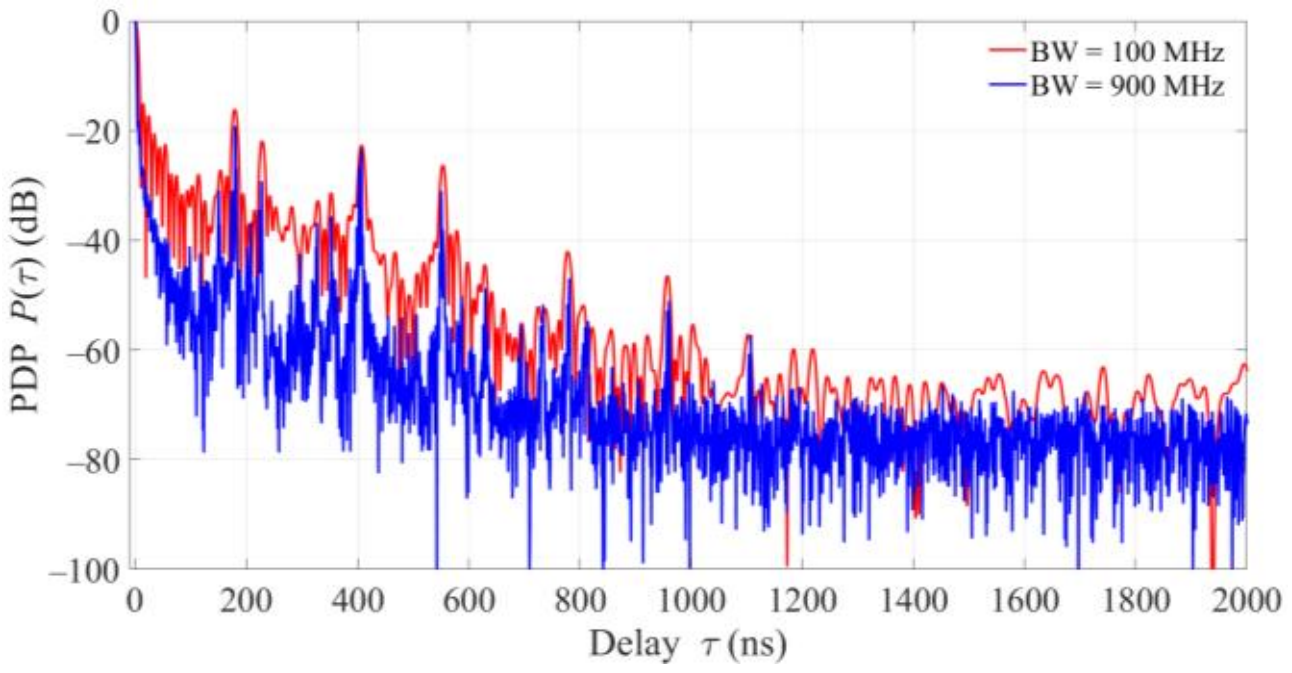

Fig. 3. Averaged PDPs for 100 and 900 MHz BWs, carrier frequency of 38 GHz, and at Tx–Rx distance of 30 m.

As the distance increases to 30 m (see Fig. 3), the overall received power decreases, and the relative level of the delayed components becomes more pronounced. The PDP shape becomes more spread in time, indicating an increase in delay spread and stronger influence of multipath reflections from corridor walls and surrounding objects. Despite this, the high-bandwidth (900 MHz) waveform still enables better separation of closely spaced echoes.

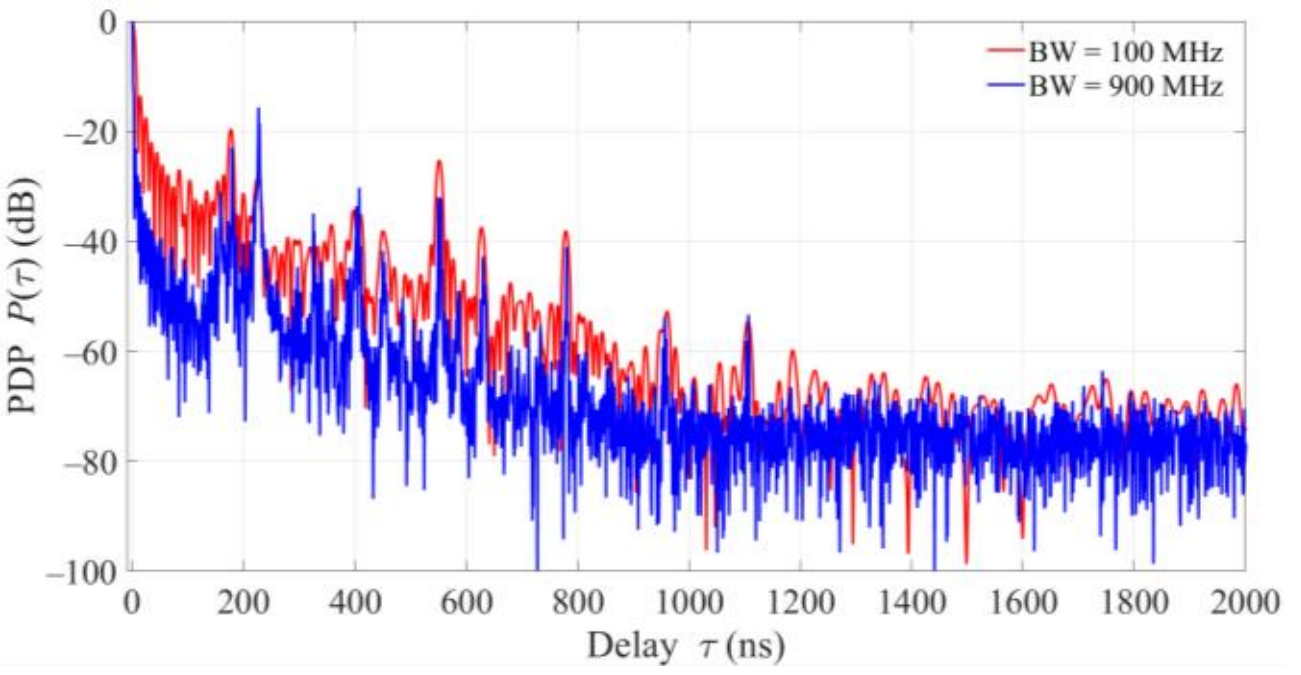

Fig. 4. Averaged PDPs for 100 and 900 MHz BWs, carrier frequency of 38 GHz, and at Tx–Rx distance of 50 m.

For the largest Tx–Rx separation equal to 50 m (see Fig. 4), the LOS component remains visible but at a significantly lower power level. This behavior confirms the growing temporal dispersion of the channel with increasing Tx–Rx distance. In addition, the reduced signal-to-noise ratio (SNR) at larger distances increases the apparent background noise level in the PDP.

Across all distances, the comparison between the two BWs demonstrates that wider BWs (i.e., 900 MHz) provide higher delay resolution, allowing a more detailed identification of multipath components and a more accurate estimation of temporal parameters such as mean delay and RMS delay spread. Conversely, the narrower 100 MHz BW produces smoother PDPs with lower temporal resolution, which can mask weaker reflections and lead to an overestimation of delay spread. In summary, the BW of the sounding signal affects the statistics and transmission characteristics of the tested channels.

Then, for the obtained PDPs, the time-dispersion parameters, i.e., the mean excess delay and RMS delay spread, were calculated based on (1) and (2), respectively. These parameters quantify the average propagation delay and the degree of temporal spreading of the received signal components, key indicators of channel selectivity and ISI potential.

Figs. 5 and 6 depict mean delay and delay spread versus Tx–Rx distance, respectively, for two analyzed BWs (i.e., 100 and 900 MHz are marked in red and blue, respectively). Additionally, regression lines for individual parameters and BWs were determined using the least squares method, which are described by the following equations

$$\bar{\tau}(D) = a_\tau \cdot D + b_\tau, \quad \sigma_\tau(D) = a_\sigma \cdot D + b_\sigma, \tag{4}$$

where $a_\tau$, $a_\sigma$ and $b_\tau$, $b_\sigma$ are the slopes (gradients) and ordinate axis-intercepts of the lines, respectively. The values of these coefficients are included in Table I.

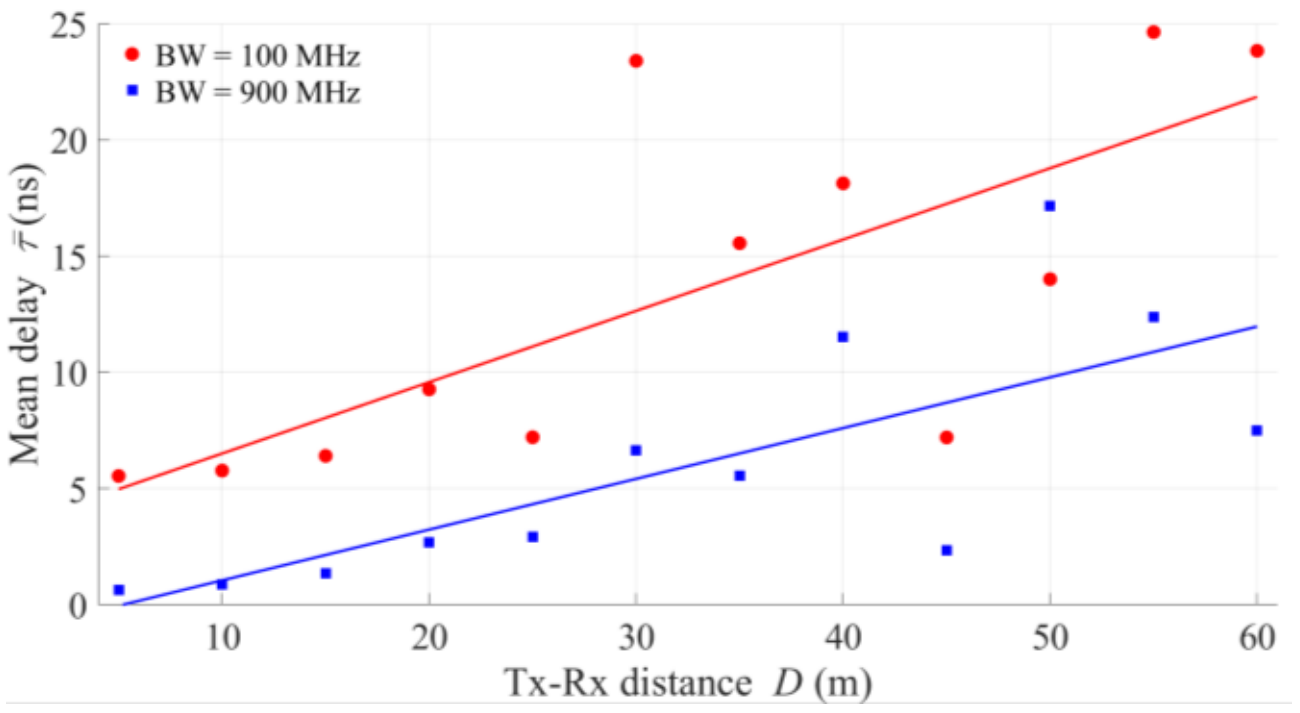

Fig. 5. Mean excess delay versus Tx–Rx distance for 100 and 900 MHz BWs, and carrier frequency of 38 GHz.

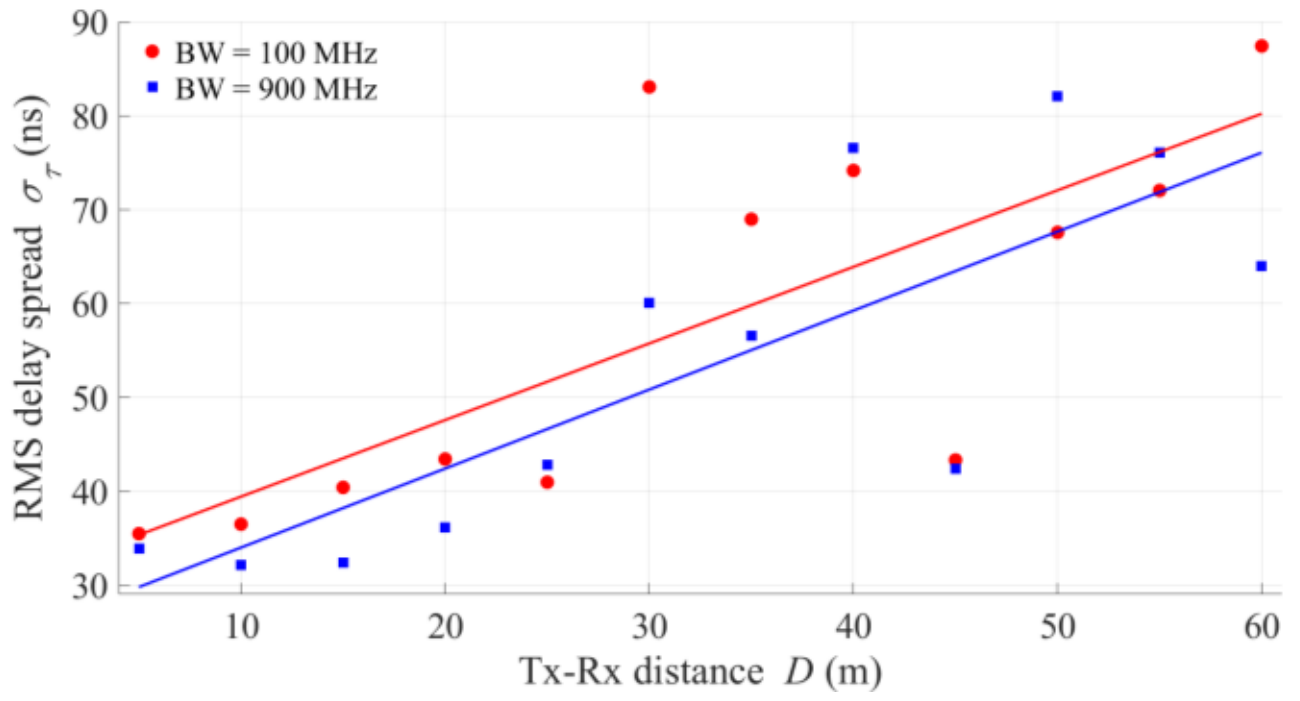

Fig. 6. RMS delay spread versus Tx–Rx distance for 100 and 900 MHz BWs, and carrier frequency of 38 GHz.

TABLE I. REGRESSION LINES' COEFFICIENTS FROM FIGS. 5 AND 6 DESCRIBED BY (4).

| **Dispersion parameters** | **BW = 100 MHz** | | **BW = 900 MHz** | |
|---|---|---|---|---|
| | $a_\tau$, $a_\sigma$ (ns/m) | $b_\tau$, $b_\sigma$ (ns) | $a_\tau$, $a_\sigma$ (ns/m) | $b_\tau$, $b_\sigma$ (ns) |
| Mean delay | 0.31 | 3.44 | 0.22 | –1.14 |
| Delay spread | 0.82 | 31.25 | 0.84 | 25.57 |

As shown, both parameters, mean delay ($\bar{\tau}$) and RMS delay spread ($\sigma_\tau$), increase approximately linearly with the distance between the Tx and Rx. This confirms that greater propagation distance is associated with longer multipath propagation paths and, consequently, larger temporal dispersion.

For both analyzed BWs, the slope coefficients $a_\tau$ and $a_\sigma$ are positive, indicating a consistent increase of delay-related parameters with distance. However, the values of these coefficients are slightly lower for the 900 MHz BW, which reflects the improved temporal resolution achieved with wider BW signals. The wider BW allows better separation of closely spaced multipath components and, therefore, results in smaller estimated mean delay and delay spread values compared to the narrower 100 MHz BW case.

The intercept values ($b_\tau$, $b_\sigma$) correspond to the expected delay parameters at minimal Tx–Rx distance. Their differences between the two BWs stem from the distinct sensitivity of the sounding signals to early reflections and noise.

Considering that the regression lines' parameters for the mean delay and RMS delay spread depend on BW, we can present them in a generalized forms

$$\begin{aligned} \bar{\tau}(D,BW) &= a_{\tau}(BW)\cdot D + b_{\tau}(BW), \\ \sigma_{\tau}(D,BW) &= a_{\sigma}(BW)\cdot D + b_{\sigma}(BW), \end{aligned} \quad (5)$$

where

$$\begin{aligned} a_{\tau}(BW) &= a_{\tau}^{a}\cdot BW + b_{\tau}^{a}, \\ b_{\tau}(BW) &= a_{\tau}^{b}\cdot BW + b_{\tau}^{b}, \\ a_{\sigma}(BW) &= a_{\sigma}^{a}\cdot BW + b_{\sigma}^{a}, \\ b_{\sigma}(BW) &= a_{\sigma}^{b}\cdot BW + b_{\sigma}^{b}. \end{aligned} \quad (6)$$

Table II draws the parameter values for the above regression lines.

TABLE II. REGRESSION LINES' COEFFICIENTS DESCRIBED BY (6).

| **Dispersion parameters** | $a_{\tau}^{a}, a_{\sigma}^{a}$ (ns/m/GHz) | $b_{\tau}^{a}, b_{\sigma}^{a}$ (ns/m) | $a_{\tau}^{b}, a_{\sigma}^{b}$ (ns/GHz) | $b_{\tau}^{b}, b_{\sigma}^{b}$ (ns) |
|---|---|---|---|---|
| Mean delay | −0.1125 | 0.3213 | −5.7250 | 4.0125 |
| Delay spread | 0.0250 | 0.8175 | −7.1000 | 31.960 |

The results summarized in Table II confirm that all regression coefficients exhibit a clear dependence on the signal BW. The negative values of coefficients $a_{\tau}^{a}$ and $a_{\sigma}^{a}$ indicate that an increase in BW leads to a reduction of both the mean delay and RMS delay spread. This trend is consistent with the improved temporal resolution provided by wideband sounding signals, which allows more accurate separation of individual multipath components and thus yields smaller apparent dispersion parameters.

Moreover, the intercept coefficients $b_{\tau}^{b}$ and $b_{\sigma}^{b}$ describe the baseline values of delay-related parameters for the reference BW and distance, providing a compact empirical model for predicting the temporal behavior of the indoor mmWave channel. Consequently, the relationships defined in (5)–(6) and parameterized in Table II enable joint characterization of the influence of both distance and signal BW on the temporal dispersion of the radio channel.

## V. SUMMARY

This paper presented the results of an experimental study on the dependence of radio channel temporal dispersion on the transmitted signal BW in an indoor mmWave environment at 38 GHz. Measurements were performed using a wideband FMCW channel sounder for two BWs, 100 and 900 MHz, and for twelve Tx–Rx distances ranging from 5 to 60 m.

From the recorded sounding signals, PDPs were derived and used to estimate the mean excess delay and RMS delay spread. The results demonstrated that both parameters increase approximately linearly with the Tx–Rx distance, confirming the expected influence of propagation path extension and multipath richness on channel temporal spreading. Additionally, a clear dependence on signal BW was observed: larger BWs provided finer delay resolution and yielded lower values of both mean delay and RMS delay spread.

Empirical regression models were proposed to describe these relationships as functions of distance and BW. The generalized model parameters presented in Tables I and II enable the joint characterization of distance- and BW-dependent behavior of temporal dispersion in indoor mmWave channels.

The obtained results may serve as a basis for developing more comprehensive statistical models. They can support designing and optimizing wideband communication systems operating in the mmWave frequency range.

## ACKNOWLEDGMENT

This work was co-funded by the Czech Science Foundation under grant no. 23-04304L, the National Science Centre, Poland, under the OPUS call in the Weave program, under research project no. 2021/43/I/ST7/03294 acronym 'MubaMilWave' and by the Military University of Technology under grant no. UGB/22-059/2025/WAT, and chip-to-startup (C2S) program no. EE-9/2/2021-R&D-E sponsored by MeitY, Government of India.